\documentclass{epl}

\title{\bf Activated processes and Inherent Structure dynamics of
       finite-size mean-field models for glasses}

\shorttitle{Activated processes ...}

\author{A. Crisanti\inst{1}\footnote{e-mail:andrea.crisanti@phys.uniroma1.it}
        and
        F. Ritort\inst{2}\footnote{e-mail:ritort@ffn.ub.es}}

\institute{
\inst{1} Dipartimento di Fisica, Universit\`a di Roma ``La Sapienza'',
	 and \\
         Istituto Nazionale Fisica della Materia, Unit\`a di Roma I \\
         P.le Aldo Moro 2, I-00185 Roma, Italy.\\
\inst{2} Physics Department, Faculty of Physics \\
         University of Barcelona, Diagonal 647, 08028 Barcelona,  Spain
        }

\rec{}{}

      \pacs{64.70.Pf}{Glass transitions}
      \pacs{75.10.Nr}{Spin-glass and other random models}
      \pacs{61.20.Gy}{Theory and models of liquid structure}
\begin{document}

\maketitle

\begin{abstract}
We investigate the inherent structure (IS) dynamics of mean-field {\it
finite-size} spin-glass models whose high-temperature dynamics is
described in the thermodynamic limit by the schematic Mode Coupling
Theory for super-cooled liquids. Near the threshold energy the
dynamics is ruled by activated processes which induce a logarithmic
slow relaxation. We show the presence of aging in both the IS
correlation and integrated response functions and check the validity
of the one-step replica symmetry breaking scenario in the presence of
activated processes. Our work shows: 1) The violation of the
fluctuation-dissipation theorem is given by the configurational
entropy, 2) The intermediate time regime ($\log(t)\sim N$) in
mean-field theory automatically includes activated processes opening
the way to analytically investigate activated processes by computing
corrections beyond mean-field.
\end{abstract}

After many years of research on the structural glass problem a lot of
experimental data has been collected but a convincing theory is still
needed \cite{ANGELL,BOU1}. The two most successful theories for the
glass transition are the Adam-Gibbs-DiMarzio and the ideal Mode Coupling
Theory (MCT) \cite{REVIEW}.  Despite their different character both are
mean-field theories. In the former case the mean-field aspect lies in
the notion of configurational entropy which assumes a breaking of the
phase space in disconnected ergodic components. In the latter the
presence of the MCT transition marks the onset of ergodicity breaking
where the diverging of a characteristic time occurs. Both approaches
have been successfully unified in the context of spin-glass theories
\cite{KTW}.

To go beyond mean-field it is necessary to include activated
processes, a very difficult task since it implies the
knowledge of the excitations involved in the dynamics. 
Recent theoretical and numerical results clearly show that the slowing down 
of the dynamics near the structural glass transition is strongly connected 
to the complex topology of the potential energy landscape \cite{GO}. 
In a glass-forming systems this is made by many deep valleys connected by 
saddles. The time evolution can then be divided into
an {\it intra-valley} and an {\it inter-valley} motion.  When the
temperature is lowered down to the order of the critical
MCT temperature $T_{\rm MCT}$ the two motions become well separated in
time and relaxation dynamics, dominated by inter-valley processes,
slows down displaying non-exponential behavior. 

To deal with this picture Stillinger and Weber (SW) \cite{SW82}
introduced the concept of {\it inherent structure} (IS) defined as the
local stable minima of the potential energy reached through a steepest
descent energy minimization process.  All configurations which under
this mapping flow to the same IS define the basin (of attraction) of
the IS. It is now a simple matter to define an IS-based thermodynamics
by replacing the partition sum with a sum over IS
\cite{SW82,SKT99}. Direct consequence of this is the introduction of a
{\it configurational entropy} $s_c(e)$ which counts the number of
different IS with the same energy $e$: $\Omega(e)=\exp(Ns_c(e))$. In
is clear that as long as the configurations retained in an IS-based
partition sum are those dominating the thermodynamics at the given
temperature $T$, the equilibrium behavior is correctly described.
However, despite of the fact that the $s_c(e)$ so defined is a {\it
dynamical quantity}, it is far from obvious that it describes the
long-time non-equilibrium behavior. The reduction from the real
dynamics to an IS dynamics is what, in the theory of dynamical
systems, is called a {\it symbolic dynamics}. This describes correctly
the dynamics only if it is associated to a {\it generating partition}
\cite{BeSc}.  In general for a generic dynamics it is not at all
trivial to demonstrate that such a partition exists, and even if it
does exist, how to find it.  Nevertheless, we can argue that in a many
valley dynamics with activated dynamics the SW mapping should be a
``good'' mapping. Indeed, since the SW mapping replaces each
configuration in a IS-basin with the IS itself, it is clear it will be
a good mapping for the long time non-equilibrium dynamics if the
system spends a lot of time inside the basins.  Under this assumption
the dynamics on time scales larger than the typical residence time
inside a IS-basin should be quite well described by the IS dynamics.
Recent numerical results on Lennard-Jones mixtures \cite{KST99}
supports this scenario for supercooled liquids. However, this mapping
may not be valid for all kind of non-equilibirum systems
\cite{MB}. Actually, for coarsening systems the SW mapping is not
meaningful \cite{CRRS}.

In this Letter we extend the analysis to {\em finite-size} mean-field
glass models and propose that activated processes seen in supercooled
liquids can be treated at a mean-field level by including {\it
finite-size effects} in the dynamics of an infinite mean-field system
going beyond the saddle-point approximation, i.e., beyond the ideal MCT.
This observation is quite reminiscent of the dynamical approach of
Sompolinsky \cite{SOMPO} and opens the way to address activated
processes in structural glasses starting from mean-field theories.  After the
work by Sompolinsky the inclusion of activated processes from this point
of view has not attracted much attention \cite{HORNER} and mean-field
dynamical studies have mainly considered the $N\to\infty$ limit before
the large-time limit \cite{REVIEW}. Other possible approaches analyze activated
process by considering instanton solutions of the mean-field equations
\cite{LI,BK}.

This work is the natural continuation of a previous one \cite{CR1} where
we introduced finite-size mean-field glasses for the analysis of the SW
configurational entropy. Here we extend our study to the long-time
non-equilibrium dynamics finding that, once finite-size effects are
included and activated processes appear in a natural way, relaxational
dynamics is driven by the configurational entropy giving support to the
one-step replica-symmetry breaking (RSB) scenario beyond mean-field.  The
present results are in agreement with recent simulations on
Lennard-Jones systems \cite{FDT} although here we go further and verify
the explicit connection between the violation of the
fluctuation-dissipation theorem and the configurational entropy.
Moreover, our results give further support to the relevance of
$p$-spin-like models for the description of the glass transition in
structural glasses.

Following Ref. \cite{CR1} we consider the dynamics of the Ising-spin Random 
Orthogonal Model (ROM) \cite{MPR94,PP95} defined by the Hamiltonian
\begin{equation}
\label{eq:ham}
  H = - 2 \sum_{ij} J_{ij}\, \sigma_i\, \sigma_j 
\end{equation}
where $\sigma_i=\pm 1$ are $N$ Ising spin variables, and $J_{ij}$ is a
$N\times N$ symmetric random orthogonal matrix with $J_{ii}=0$.  We
use the heat-bath Monte Carlo scheme with random sequential spin
updating.  This model presents a thermodynamic glass transition typical of
mean field $p$-spin glasses with $p>2$ \cite{CHS93} 
at $T_{\rm MCT}=0.536$ with threshold energy per spin 
$e_{\rm th} = -1.87$.  
Below this temperature the system is dynamically confined
into a metastable state (basin or valley) and cannot reach true
equilibrium. A equilibrium transition with collapse of the
configurational entropy takes place below $T_{\rm MCT}$ at the Kauzmann
temperature $T_{\rm c}=0.25$, with critical energy per spin 
$e_{\rm c}= -1.936$ \cite{MPR94,PP95}. 
The analysis of the free-energy landscape of this model
\cite{PP95} reveals that the phase space is composed by an
exponentially large (in $N$) number of different basins, separated by
infinitely large (for $N\to\infty$) barriers. Above $T_{\rm MCT}$ and in the
large $N$ limit the IS with $e=e_{\rm th}$ attract most (exponentially in
$N$) of the states and dominate the behavior of the system. For finite
$N$ basins of IS with $e \not= e_{\rm th}$ have statistical weight and
may influence the dynamics \cite{CR1}.

To study the non-equilibrium relaxational dynamics we quench 
the system at time $t=0$ from an
equilibrium state at temperature $T_{\rm i}>T_{\rm g}$ to a final
temperature $T_{\rm f}<T_{\rm g}$. The glass transition temperature
$T_{\rm g}$ is defined, in accordance with the ``experimental'' definition,
 as the temperature below which we cannot equilibrate the system on the 
longest Monte Carlo run. The associated IS dynamics is obtained 
by regularly quenching the system down to $T=0$ from the relaxing
configuration and recording the IS associated with the
instantaneous basin.  
We shall consider both one-time quantities, as the average IS energy, 
and two-time quantities, as correlation and response functions.

\begin{figure}
\onefigure{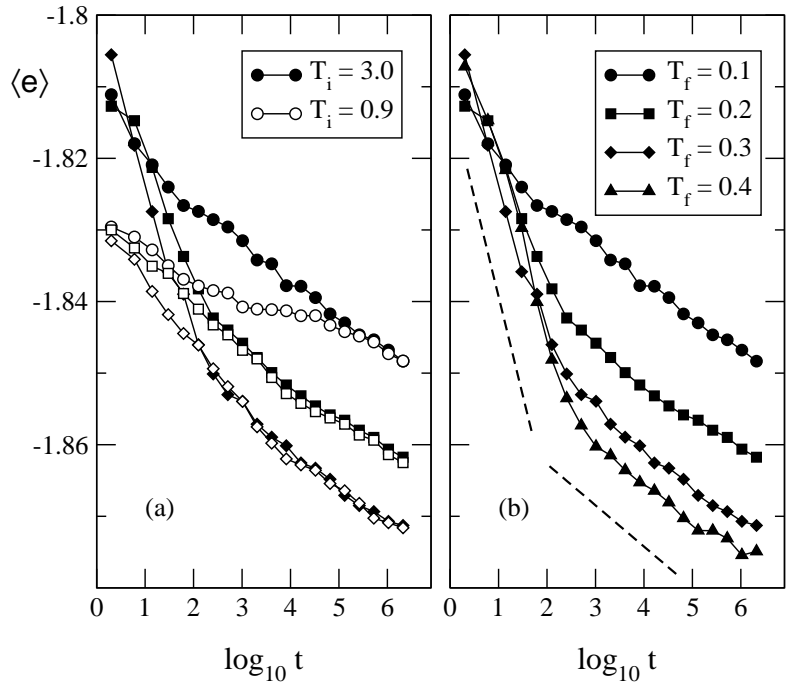}
\caption{Average IS as a function of time.
	 $(a)$: Filled symbols $T_{\rm i} = 3.0$, 
                empty symbols $T_{\rm i} = 0.9$. Final temperatures are as
                in panel $(b)$.
	 $(b)$: $T_{\rm i} = 3.0$.
         The average is over different equilibrium initial configurations.
}
\label{fig:f1}
\end{figure}

{\em Relaxation of one-time quantities.}  In figure \ref{fig:f1} it is
shown the average IS energy per spin $\langle e\rangle(t)$ as function
of time for a system with $N=300$ spins.  The analysis of the figure
reveals that the relaxation process can be divided into two different
regimes where $\langle e\rangle(t)$ decreases with different power laws.
The slope of the decay in the first regime is independent of $T_{\rm f}$
while the slope in the second regime is independent of both $T_{\rm i}$
and $T_{\rm f}$ [cfr. panel $(a)$]. For a given $T_{\rm i}$ the final
temperature $T_{\rm f}$ only controls the cross-over between the two
regimes. A similar behavior has been observed in molecular dynamics
simulations of supercooled liquids \cite{KST99}.  We note that since we
use discrete variables, and hence a faster dynamics, the very-early
regime observed in \cite{KST99} where $\langle e\rangle(t)$ is almost
independent of $t$ is absent. Moreover since we use fully connected
systems the power law exponents change with $N$, even if the qualitative
scenario is unchanged.  The two regimes are associated with different
relaxation processes.  In the first part the system has enough energy
and relaxation is mainly due to {\it path search} out of basins through
saddles of energy lower than $k_{\rm B}T_{\rm f}$.  During this process
the system explores deeper and deeper valleys while decreasing its
energy. The process stops when all barrier heights become of $O(k_{\rm
B}T_{\rm f})$. From now on the relaxation can only proceed via activated
processes and the dynamics slows down becoming logarithmic in time. Note
that, for finite $N$, the activated regime starts already above the
threshold indicating that not only stable states exist above $e_{\rm
th}$ but these also influence the hopping dynamics.

\begin{figure}
\onefigure{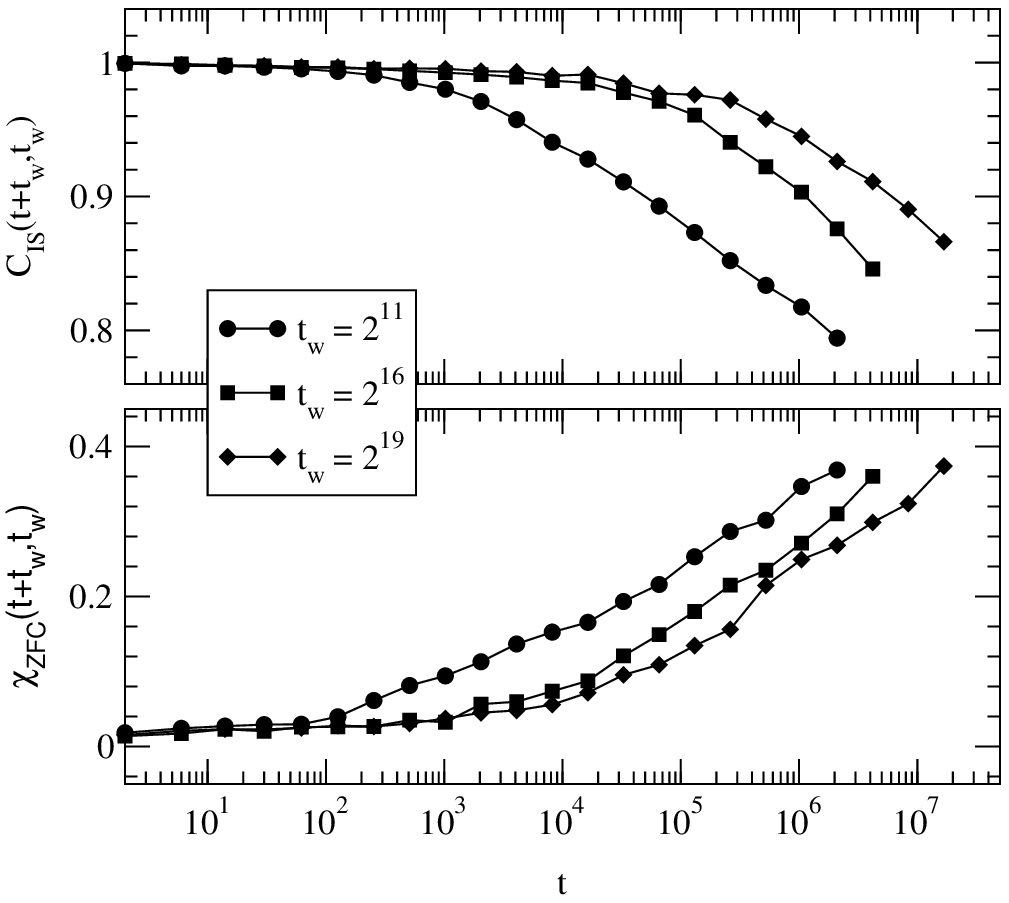}
\caption{IS correlation (above) and integrated response (below) functions
         as function of time for different waiting times.
         The system size is $N=300$, $T_{\rm i}=3$, $T_{\rm f}=0.2$ 
         and $T_{\rm g}\simeq 0.5$.
         Data have been averaged over about $400$ dynamical histories. 
}
\label{fig:f2}
\end{figure}

{\em Aging in correlation and response functions.} More information on the
non-equilibrium activated regime can be obtained from the analysis of
correlation and response functions. In order to study the relevance of the SW
mapping we consider the IS-based correlation and response functions 
defined as,
\begin{equation}
C_{\rm IS}(t,s)=\frac{1}{N}\sum_{i=1}^N\,\sigma^0_i(t)\,\sigma^0_i(s)
\label{eqC}
\end{equation}
\begin{equation}
R_{\rm IS}(t,s)=\frac{1}{N}\sum_{i=1}^N\,\frac{\delta\sigma^0_i(t)}{\delta
h_i(s)} \qquad t>s
\label{eqR} 
\end{equation} 
where $\sigma^0(t)$ is the IS at time $t$ and $h_i$ an
external field. It can be shown that in equilibrium the 
fluctuation-dissipation theorem $TR_{\rm IS}(t-s)=-\partial_t C_{\rm IS}(t-s)$ 
holds \cite{CR2}.  To study the non-equilibrium response function we
quench the system at time $t=0$ from an equilibrium state at 
$T_{\rm i}>T_{\rm g}$  to $T_{\rm f}<T_{\rm g}$ and, 
after a given waiting time $t_{\rm w}$,  
we make a {\it clone} of it to which a (small) constant uniform 
magnetic field is applied. 
To reduce fluctuations the same stochastic noise is used for both
replicas. The experiment is repeated for different (small) fields to
control the linear response regime.
Results for the correlation function and the integrated 
response, or zero-field cooled susceptibility,
$\chi_{\rm ZFC}(t,t_{\rm w})=
       \int_{t_{\rm w}}^tdt'R_{\rm IS}(t,t')$ are reported 
in figure \ref{fig:f2}. The presence of aging is rather
clear.

{\em The one-step RSB scenario and the SW configurational entropy.}  A
way to see how equipartition is broken in the non-equilibrium regime is
through the fluctuation-dissipation ratio, initially introduced in the
context of spherical models for spin glasses
\cite{REVIEW,CHS93,CUKU,FRME}, which gives a direct measure of frozen
degrees of freedom in the response of the system. Usually the broken of
equipartition is expressed through an effective temperature defined as
$T_{\rm eff}(t,t_{\rm w})=\partial_{t_{\rm w}} C_{\rm IS}(t,t_{\rm w})
/R_{\rm IS}(t,t_{\rm w})$, or alternatively as
\begin{equation} 
T_{\rm eff}^{-1}(C_{\rm IS}) =
        -\frac{\partial \chi_{\rm ZFC}(t,t_{\rm w})}
              {\partial C_{\rm IS}(t,t_{\rm w})}.
\label{eqTeff}
\end{equation}
which is the slope of the curves $\chi_{\rm ZFC}$ versus $C_{\rm IS}$.
This temperature reduces to $T$ in equilibrium and is larger 
when equipartition is broken. 

In the context of mean-field theories for the glass transition 
$T_{\rm eff}$ is given by the constant temperature derivative of 
the free-energy with respect  to the configurational entropy: 
$T_{\rm eff}^{-1}=[\partial s_c(f) / \partial f]_T$
\cite{MON,MEZARD}, which in the present IS analysis means
\begin{equation}
 T_{\rm eff}^{-1} = \left[\frac{\partial s_{c}(f_{\rm IS})}
                               {\partial f_{\rm IS}}
                    \right]_T  
                  = \left[\frac{\partial s_c(e)}{\partial e}\right]_T 
			\Big\slash
                    \left[\frac{\partial f_{\rm IS}(e)}{\partial e}\right]_T 
\label{sc}
\end{equation}
where $f_{\rm IS}$ is the free energy of the IS with energy $e$, 
i.e., obtained from the partition sum {\it restricted} to the 
configurations in the IS-basins of IS with energy $e$.
Writing $f_{\rm is} = e + \Delta e(T,e) - T\,s_{\rm IS}(T,e)$ 
we see that $[\partial f_{\rm IS}(e) /\partial e]_T = 1 + \delta(e,T)$,
where the last term measure how IS with different $e$ differ one each other.

\begin{figure}
\onefigure{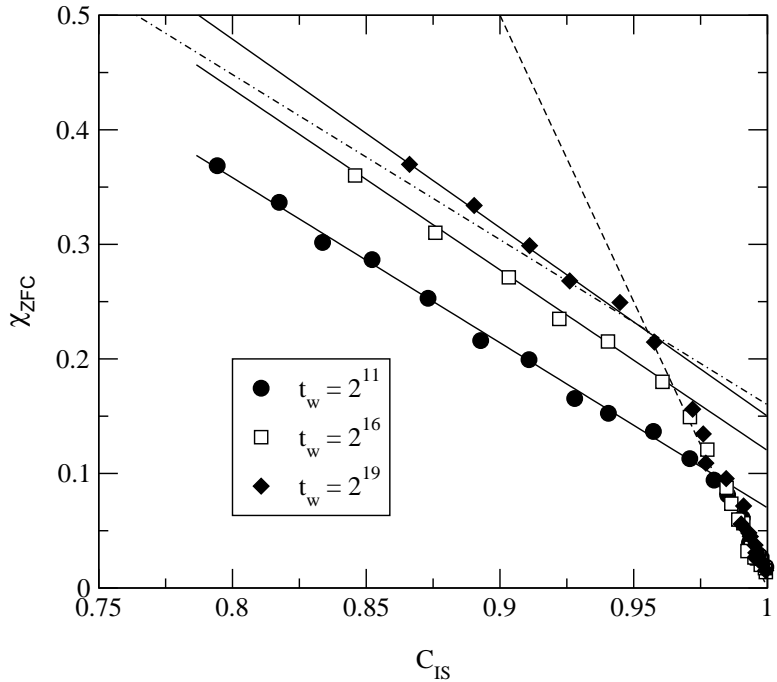}
\caption{Integrated response function as a function of IS correlation
function.  The data are from Fig. \protect\ref{fig:f2}.  The dash line
has slope $\beta_{\rm f} = 5.0$, while the full lines is the prediction
(\protect\ref{sc}) with $\delta(e,T)=0$ and $s_c(e)$ from
Ref. \protect\cite{CR1}: $T_{\rm eff}(2^{11})\simeq 0.694$, 
$T_{\rm eff}(2^{16})\simeq 0.634$ and
$T_{\rm eff}(2^{19})\simeq 0.608$.
The dot-dashed line is $\beta_{\rm eff}$ for
$t_{\rm w}=2^{11}$ drawn for comparison.  }
\label{fig:f3}
\end{figure}

Results from our numerical simulations are shown in figure \ref{fig:f3}. 
For each $t_{\rm w}$  two distinct behaviors are clearly seen. 
For large values of $C_{\rm IS}$  the slope of all the curves is
$\beta_{\rm f}=1/T_{\rm f}$.  Below a $t_{\rm w}$-dependent 
value of $C_{IS}$ the curve changes slope and the effective 
temperature increases. In order to compare our results with prediction
(\ref{sc}), we have to compute $f_{\rm IS}$. This can be done from the
knowledge of the probability that an equilibrium configuration at temperature 
$T=1/\beta$ lies in a basin associated with IS of energy between $e$ 
and $e+de$:
$P_N(e,T) =  \exp\, N\,[-\beta f_{\rm IS} + s_c(e) + \beta f(T)]$
where $f(T)$ is the full free-energy. Using the results from 
Refs. \cite{PP95,CR1} both $f_{\rm IS}$ and its derivative can be estimated.
In all cases we found $\delta(e,T)\simeq 0$, within numerical errors, a 
results in agreement with the fact that the ROM is rather similar to 
the Random Energy Model \cite{REVIEW} for which $\delta(e,T)=0$. 
Using for $s_c(e)$ the expression obtained in Ref. \cite{CR1} we obtain
from (\ref{sc}) 
the slopes $\beta_{\rm eff} = 1/T_{\rm eff}$ shown in figure \ref{fig:f3}.
The agreement is rather good.
Data for $T_{\rm f}=0.1$ and $0.3$ are more noisy, but consistent with this
identification. 

We therefore conclude that in the activated regime the effective
temperature $T_{\rm eff}$ derived from the SW configurational entropy
agree extremely well with the numerical data confirming the one-step
scenario. This is a highly non-trivial result.  First of all, as
discussed above, it is not obvious that the SW mapping correctly
describes the long-time non-equilibrium dynamics.  It is plausible that
it works for an activated dynamics, but plausibility is not a proof.
Second, equation (\ref{sc}) is derived within mean-field, i.e., for
$N\to\infty$. Here we apply it for states which exist only for {\it
finite} $N$ being, as discussed above, those which govern the
relaxational dynamics in this regime but disappear in the thermodynamic
limit. Thus our finding gives to (\ref{sc}) a broader validity.  We note
that quite new results from numerical simulations of Lennard-Jones
mixtures are in agreement with (\ref{sc}), but with $\delta(e,T)\not=0$
\cite{Sciort}.

 From the discussion on the SW approach it follows that if
the SW mapping defines a good mapping then it should equally well describe 
the long-time relaxational dynamics regardless which configuration we use
to identify the IS-basin. To check this point we
have repeated the simulations using slightly different definitions of IS, 
i.e., changing the minimization rule. In all 
cases we have found the same results, so
that in this case the SW mapping provides a 
good ``symbolic'' dynamics for the long-time non-equilibrium behavior.

We now summarize our findings. The Stillinger-Weber decomposition of
phase space in terms of IS is a natural and simple statistical
description of a dynamical system with activated dynamics among
different basins when the time scales for motions inside a basin and
between basins are well separated. Moreover we have shown that
finite-size mean-field glasses are valuable models to describe activated
processes in structural glasses.  In the activated regime these models
display slow logarithmic relaxation with two class of motions which
confirm the one-step replica symmetry breaking scenario: a fast
intra-basin motion where fluctuation-dissipation holds and a slow
inter-basin motion corresponding to activated processes.  The
fluctuation-dissipation ratio in the activated regime can be fully
described in terms of the configurational entropy
generalizing the mean-field scenario with the inclusion of activated
processes. Our results suggest the possibility of investigating the glass
transition in structural glasses from finite-size correction to
mean-field theory justifying efforts in this direction.

\acknowledgments We thank for useful discussions C. Donati,
S. Franz, F. Sciortino and A. Rocco for a careful reading of the manuscript. 
F.R is supported by the Ministerio de Educaci\'on y Ciencia in Spain
(PB97-0971).

\vskip -0.5truecm
%%%%%%%%%%%%%%%%%%%%%%%%%% REFERENCES %%%%%%%%%%%%%%%%%%%%%

%%%%%%%%%%%%%%%%%%%%%%%%%% FIGS  %%%%%%%%%%%%%%%%%%%%%
\end{document}